\documentclass[aps,jcp,preprint,a4paper,unsortedaddress,onecolumn]{revtex4}

\usepackage[fleqn]{amsmath}
\usepackage{amssymb}
\usepackage[dvips]{graphicx}
\usepackage{color}
\usepackage{tabularx}
\usepackage{mathtools}
\usepackage{algpseudocode}
\usepackage{enumitem}


\newcommand{\vect}[1]{\textbf{\textit{#1}}}

\newcommand{\dir}{\mathrm{dir}}
\newcommand{\rec}{\mathrm{rec}}
\newcommand{\corr}{\mathrm{correction}}
\newcommand{\correlation}{\mathrm{corr}}
\newcommand{\erfc}{\mathrm{erfc}}

\newcommand{\ik}{\mathrm{ik}}
\newcommand{\ad}{\mathrm{ad}}
\newcommand{\homo}{\mathrm{homo}}
\newcommand{\hetero}{\mathrm{inhomo}}
\newcommand{\sinc}{\mathrm{sinc}}
\newcommand{\self}{\mathrm{self}}

\newcommand{\suma}{\sum_{\alpha}}

\newcommand{\sumll}{\sum_{l\neq 0}}

\newcommand{\sumlab}{\sum_{l_1,l_2\neq 0}}
\newcommand{\za}[1]{Z_{\alpha,l}(\vect #1)}
\newcommand{\zaa}[1]{Z_{\alpha,l_1}(\vect #1)}
\newcommand{\zbb}[1]{Z_{\beta,l_2}(\vect #1)}

\newcommand{\qh}{q_{\mathrm{H}}}
\newcommand{\qo}{q_{\mathrm{O}}}
\newcommand{\sh}{\vect s_{\mathrm{H}}}
\newcommand{\so}{\vect s_{\mathrm{O}}}
\newcommand{\myphi}{\varphi_{n}}
\newcommand{\hmyphi}{\hat{\varphi}_{n}}

\newcommand{\mo}{\mathcal {O}}
\newcommand{\me}{\mathcal {E}}
\newcommand{\mg}{\boldsymbol{\mathcal {G}}}
\newcommand{\mf}{\boldsymbol{\mathcal {F}}}

\newcommand{\mga}[1]{\mg_{\alpha,l}(\vect #1)}
\newcommand{\mgab}[1]{\mg_{\alpha,l_1;\beta,l_2}(\vect #1)}
\newcommand{\mfa}[1]{\mf_{\alpha,l}(\vect #1)}
\newcommand{\mfab}[1]{\mf_{\alpha,l_1;\beta,l_2}(\vect #1)}

\begin{document}

\title{Multiple Staggered Mesh Ewald: Boosting the Accuracy of the Smooth Particle Mesh Ewald Method}
\author{Han Wang}
\email{wang_han@iapcm.ac.cn}
\affiliation{Institute of Applied Physics and Computational Mathematics, Fenghao East Road 2, Beijing 100094, P.R.~China}
\affiliation{CAEP Software Center for High Performance Numerical Simulation, Huayuan Road 6, Beijing 100088, P.R.~China}
\author{Xingyu Gao}
\affiliation{Laboratory of Computational Physics, Huayuan Road 6, Beijing 100088, P.R.~China}
\affiliation{Institute of Applied Physics and Computational Mathematics, Fenghao East Road 2, Beijing 100094, P.R.~China}
\affiliation{CAEP Software Center for High Performance Numerical Simulation, Huayuan Road 6, Beijing 100088, P.R.~China}
\author{Jun Fang}
\affiliation{Institute of Applied Physics and Computational Mathematics, Fenghao East Road 2, Beijing 100094, P.R.~China}
\affiliation{CAEP Software Center for High Performance Numerical Simulation, Huayuan Road 6, Beijing 100088, P.R.~China}
   
\begin{abstract}
  The smooth particle mesh Ewald (SPME) method 
  is the standard method for computing the electrostatic interactions in the molecular simulations.
  In this work, the multiple staggered mesh Ewald (MSME) method is proposed to
  boost the accuracy of the SPME method.
  Unlike the SPME that achieves higher accuracy by refining the mesh,
  the MSME improves the accuracy
  by averaging the standard SPME forces computed on, e.g.~$M$, staggered meshes.
  We prove, from theoretical perspective, that
  the MSME is as accurate as the SPME,
  but uses $M^2$ times less mesh points in a certain parameter range.
  In the complementary parameter range, the MSME is as accurate as the SPME with
  twice of the interpolation order.
  The theoretical conclusions are numerically validated both
  by a uniform and uncorrelated charge system,
  and by a three-point-charge water system that is widely used as solvent for
  the bio-macromolecules.\\
  

\noindent\textbf{Keywords}: Molecular simulation, electrostatic interaction, smooth particle mesh Ewald method, multiple staggered mesh Ewald  
\end{abstract}

\maketitle

\section{Introduction}
The electrostatic interaction is one of the most important interactions, and, perhaps,
the most computationally demanding part in molecular dynamics (MD) simulations.
One popular way of computing the electrostatic interaction is the Ewald summation~\cite{ewald1921die}.
It was shown that
the optimal computational complexity of this method is 
$\mo(N^{3/2})$~\cite{perram1988asc} ($N$ being the number of point charges in the system),
therefore, as the number of charges increases, e.g.~to several hundreds~\cite{pollock1996comments},
the Ewald summation becomes relatively expensive.
This stimulates the development of
the Ewald-based fast algorithms like the particle mesh Ewald (PME) method~\cite{darden1993pme}, the smooth particle mesh Ewald (SPME) method~\cite{essmann1995spm}
the particle-particle particle-mesh (PPPM) method~\cite{deserno1998mue1} and
the methods based on the non-equispaced fast Fourier transforms~\cite{pippig2013pfft}.
All these fast methods reduce the  
computational complexity to $\mo (N\log N)$
by interpolating the charges on a mesh and 
solving the Poisson's equation with the fast Fourier transform (FFT). 
The Ewald-based fast methods are also proposed to compute the dispersion interactions
in the systems that present interfaces~\cite{alejandre2010surface,isele2012development,ismail2007application}.
They are shown to be more accurate and even faster than
the traditional treatment of the dispersion interactions in these systems, viz.~using a large cut-off radius~\cite{isele2013reconsidering,wennberg2013lennard}.

The main difficulty of applying the Ewald-based fast methods in practical simulations
is how to determine the working parameters.
It is well known that the arbitrarily chosen parameters may lead to a substantial slowdown of the computation,
or results that are several orders of magnitudes less accurate.
This problem may be solved, in \emph{a posteriori} manner, by
comparing the computed forces while scanning the parameter space for a representative snapshot of the system~\cite{abraham2011optimization}.
An alternative way is by a parameter tuning algorithm that
automatically determines the most efficient combination of parameters under the restraint of a prerequisite accuracy~\cite{wang2010optimizing}.
The success of this algorithm relies on the quality of the \emph{a priori} error estimate
that quantitatively describes the accuracy of the fast methods as a function of working parameters,
and a large amount of work have been dedicated to this direction~\cite{neelov2010interlaced,kolafa1992cutoff,hummer1995numerical,petersen1995accuracy,deserno1998mue2,stern2008mesh,wang2010optimizing,wang2012error,wang2012numerical}.

When tuning the parameters,
the standard way to increase the accuracy (in the reciprocal space)
is to refine the FFT mesh or to use interpolation basis of higher orders.
Recently, a non-standard way, i.e.~the staggered meshes,
was introduced in the SPME and PPPM,
and the new methods are called the staggered mesh Ewald (SME)~\cite{cerutti2009staggered}
and the interlaced PPPM method~\cite{neelov2010interlaced}, respectively.
The SME is theoretically proved to be always more accurate than its non-staggered mesh counterpart~\cite{wang2012numerical}.

In this work, we develop the multiple staggered mesh Ewald (MSME) method.
It takes the average of the reciprocal forces computed by the SPME method on $M$ meshes,
which are shifted to the $M$ equally partitioning points of the mesh subcell diagonal.
The MSME method uses $M$ times more FFT mesh points,
and achieves, in a certain parameter range,
the same accuracy that would need $M^3$ times more FFT mesh points in the standard SPME.
In the complementary range of parameters, MSME achieves the same accuracy
as the SPME that uses twice of the interpolation order.
These properties of the MSME are proved, from theoretical perspective, by a systematical error estimate,
in which the accuracy is described as a function of the working parameters like the mesh and the interpolation order.
The quality of the error estimate is numerically validated
both by a uniform and uncorrelated point charge system,
and by a rigid three-point-charge water model that is widely used as solvent for
biological macromolecules.



This paper is organized as follows. In Section~\ref{sec:ewald}, the Ewald summation and the particle mesh Ewald method are briefly introduced.
The notations are setup and the difference between our implementation of the SPME and that proposed in the original paper is pointed out.
In Section~\ref{sec:msme}, the MSME method is introduced, and its accuracy is numerically discussed and compared to the SPME method.
In Section~\ref{sec:error}, 
the numerical phenomena of MSME is analyzed by the error estimate.
Section~\ref{sec:error-numeric} validates the quality of the error estimate by comparing it to the numerically computed error.
In Section~\ref{sec:conclusion}, the work is concluded and 
the performance issue is discussed in more details.

\section{The Ewald summation and the particle mesh Ewald method}
\label{sec:ewald}
We denote the $N$ charged particles in the system
by $\{q_1, \cdots, q_N \}$, and their position by
$\{\vect r_1, \cdots, \vect r_N\}$.
If the system is subject to the periodic boundary condition,
then the electrostatic interaction is given by:
\begin{align}\label{eqn:ele}
  E = \frac12 \sum_{\vect n}^\ast\sum_{i, j = 1 }^N
  \frac{q_i q_j}{\vert \vect r_{ij} + \vect n\vert},
\end{align}
where $\vect r_{ij} = \vect r_i - \vect r_j$.
$\vect n = n_1\vect a_1 + n_2\vect a_2 + n_3\vect a_3$
is the lattice 
with $(n_1, n_2, n_3)\in \mathbb Z^3$ and $(\vect a_1, \vect a_2, \vect a_3)$
are the unit cell vectors.
When $\vect n = 0$,
the inner summation runs over all Coulomb interactions
between the particle pairs in the unit cell,
otherwise,
it is counting the interaction between the unit cell and its periodic images.
The ``$\ast$'' over the outer summation means
that when $\vect n = 0$, $i \neq j$. 

The Ewald summation decomposes the electrostatic interaction~\eqref{eqn:ele}
into three
parts, the direct part, the reciprocal part and the correction
part
\begin{equation}\label{eqn:es}
E = E_{\dir} + E_{\rec} + E_{\corr},
\end{equation}
with the definitions
\begin {align}\label{eqn:es-dir}
  E_{\dir}
  & =
  \frac12 \sum^{\ast}_{\vect n}\sum_{i,j = 1}^{N}
  \frac{q_iq_j \erfc(\beta \vert\vect{r}_{ij} + \vect{n}\vert)}
  {\vert\vect{r}_{ij} + \vect{n}\vert},
  \\\label{eqn:es-rec}
  E_{\rec}
  & =
  \frac1{2\pi V} \sum_{\vect m \neq 0}
  \frac{\exp(-\pi^2\vect m^2 / \beta^2)}{\vect m^2}
  S(\vect m) S(-\vect m), \\\label{eqn:es-cor}
  E_{\corr}
  & =
  -\frac\beta{\sqrt \pi} \sum_{i=1}^N q_i^2,
\end {align}
where $\vect m = m_1\vect a_1^\ast + m_2\vect a_2^\ast + m_3\vect a_3^\ast$
is the reciprocal space lattice with $(m_1, m_2, m_3)\in \mathbb Z^3$ and
$\vect a_\alpha^\ast, \ \alpha = 1,2,3$ are the conjugate reciprocal vectors
that are defined by relations
$\vect a_\alpha\cdot \vect a_\beta^\ast = \delta_{\alpha\beta}$, $\alpha,\beta = 1,2,3$.
The direction index ``$\beta$'' here should be clearly distinguished with the damping parameter $\beta$ in Eqs.~\eqref{eqn:es-dir}--\eqref{eqn:es-cor} by the context.
$V$ is the volume of the unit cell calculated by
$V = \vect a_1 \cdot(\vect a_2\times\vect a_3)$.
The $\erfc (x)$
in Eq.~\eqref{eqn:es-dir} is the complementary error function, and 
the $S(\vect m)$ in Eq.~\eqref{eqn:es-rec} is the structure factor defined by
\begin{align}\label{eqn:sf}
  S(\vect m) = \sum_{j=1}^N q_j e^{2\pi i \vect m\cdot \vect r_j}.
\end{align}
The ``$i$'' at the exponent is the imaginary unit
that should not be confused with the particle index $i$.

The complementary error function in the direct energy~\eqref{eqn:es-dir}
converges exponentially fast w.r.t.~increasing distance
$\vert \vect r_{ij} + \vect n\vert $,
thus it can be cut-off at a certain radius $r_c$.
As a consequence,
one needs to sum a finite number of terms in
\eqref{eqn:es-dir}, 
and the cost of this computation is $\mo(N)$
by using the standard neighbor list algorithm~\cite{frenkel2001understanding}.
The summation in the reciprocal energy~\eqref{eqn:es-rec} also converges
exponentially fast,
so it can be truncated at the reciprocal space cut-off: $-K_\alpha/2 \leq m_\alpha < K_\alpha/2$.
To achieve a prerequisite accuracy,
the number of Fourier modes in summation~\eqref{eqn:es-rec} should be taken the same order as the number of
charged particles in the system, i.e.~$K_1 K_2 K_3 \sim N $.
Therefore, the computational complexity of the reciprocal interaction is $\mo(N^2)$.


We temporally assume that $N = K_1 K_2 K_3$, 
and that the charged particles locate on a $K_1\times K_2\times K_3$
mesh,
then the structure factor~\eqref{eqn:sf} is nothing but a discrete Fourier transform and 
can be computed at the cost of $\mo(N \log N)$ by
using the FFT.
In MD simulations, the particles do not necessarily locate on a mesh,
so the fast methods interpolate the particle charges
on the uniform mesh points,
then use FFT to accelerate the computation.
The computational complexity of the interpolation and the FFT are
$\mo(N)$ and $\mo(N\log N)$, respectively.
The total computational expense of the fast methods is $\mo(N\log N)$.

Before developing the MSME method,
we briefly introduce the SPME method and refer the readers to Ref.~\cite{essmann1995spm} for
more details.
We let the number of mesh points on three directions be
$K_\alpha, \ \alpha = 1,2,3$. For the coordinate of a particle $\vect r$,
we introduce the notation
$
  u_\alpha = K_\alpha \vect a_\alpha^\ast \cdot \vect r
$,
and it is clear that $0\leq u_\alpha < K_\alpha$.
We approximate the complex exponential
$e^{2\pi i m_\alpha u_\alpha/K_\alpha}$ measured on particle position $u_\alpha$
by a linear combination of the complex exponentials measured on mesh (see Eq.~(S15) in the supplementary material):
\begin{align}\label{eqn:approx-exp-1d}
  e^{2\pi i m_\alpha u_\alpha/K_\alpha} \
  \approx
  \frac1{K_\alpha}\sum_{l_\alpha\in I_{K}} \frac{1}{\hmyphi(m_\alpha)} \myphi (u_\alpha -l_\alpha)
  \,e^{2\pi i m_\alpha l_\alpha/K_\alpha},
\end{align}
where $I_K = \{ \: l\: \vert \: l\in \mathbb Z, -K/2\leq l < K/2\}$.
$\myphi$ denotes the $n$-th order cardinal B-spline that is defined by the recursive formula:
\begin{align}
  \varphi_{1} (x) &= \chi_{[-\frac12,\frac12]}(x), \quad  \myphi(x) = K_\alpha \varphi_{n-1} \ast \varphi_{1} (x),
\end{align}
where $\chi_{[-\frac12,\frac12]}$ denotes the characteristic function on interval $[-\frac12,\frac12]$.
$\hmyphi(m_\alpha )$ is the Fourier transform of $\myphi$, given by
\begin{align}\label{eqn:hat-phi}
  \hmyphi(m_\alpha )  =  \frac1{K_\alpha} \Big[\sinc( \frac{\pi m_\alpha }{K_\alpha} )\Big]^n.
\end{align}
By using the approximation~\eqref{eqn:approx-exp-1d} in three dimensional space,
we reach
\begin{align}\label{eqn:approx-exp}
  e^{2\pi i \vect m\cdot \vect r}
  & \approx
  \frac{B(\vect m)}{K_1K_2K_3}
  \Big(
  \sum_{l_1,l_2,l_3}
  P_{\vect r}(l_1,l_2,l_3)
  \exp[2\pi i(\frac{m_1l_1}{K_1} + \frac{m_2l_2}{K_2} + \frac{m_3l_3}{K_3}) ]
  \Big),
\end{align}
where
\begin{align} \label{eqn:b}
  &B(\vect m) = \prod_{\alpha} \frac 1{\hmyphi (m_\alpha)}, \\
  &P_{\vect r}(l_1,l_2,l_3) = \prod_\alpha \myphi(u_\alpha - l_\alpha).
\end{align}
It should be noted here that the function $B$ defined by Eq.~\eqref{eqn:b}
should be distinguished with that defined in the original SPME paper~\cite{essmann1995spm},
saying
\begin{align}\label{eqn:b-orig}
  B(\vect m)
  =
  \prod_{\alpha}
  \Big\{
  e^{2\pi i (n-1) m_\alpha/K_\alpha}
  \Big[
  \sum_{k=0}^{n-2}
  \myphi (k+1) e^{2\pi i m_\alpha k/K_\alpha}
  \Big]^{-1}
  \Big\}.
\end{align}
Inserting the interpolation~\eqref{eqn:approx-exp} into
Eq.~\eqref{eqn:es-rec}, we have the fast method of computing
the reciprocal energy,
\begin{align}\label{eqn:approx-rec-ener}
  E_\rec
  \approx \ &  
  \sum_{l_1,l_2,l_3}
  Q(l_1,l_2,l_3)
  [\,Q \ast (F B^2)^{\vee}] (l_1,l_2,l_3),
\end{align}
where
\begin{align}
  F(\vect m)
  &=
  \frac1{2\pi V}\times
    \begin{dcases}
      \frac{\exp(-\pi^2\vect m^2 / \beta^2)}{\vect m^2} &  \vert \vect m \vert \neq 0, \\
      \: 0 & \vert \vect m\vert  = 0,
    \end{dcases}
\end{align}
and $Q$ is the charge distribution defined on the mesh
\begin{align}
  Q (l_1,l_2,l_3)= \sum_j q_j P_{\vect r_j} (l_1,l_2,l_3).
\end{align}
The symbol ``$\vee$'' denotes the backward Fourier transform.
The operator ``$\ast$'' denotes the convolution,
which can be computed by
$Q\ast (FB^2)^\vee = [ \,\hat Q\times (FB^2)\, ]^\vee$ with ``$\wedge$'' denoting the forward Fourier transform.
The r.h.s.~of the equation is roughly explained as firstly converting the
charge distribution to the reciprocal space ($\hat Q$), 
solving the Poisson's equation with smeared source ($\hat Q \times F$),
then transforming the result back ($[ \,\hat Q\times (FB^2)\, ]^\vee$).
The function $B$ appears
due to the interpolation of the point charge distribution.
By using the FFT, the computational expense of the convolution is of order
$\mo(N\log N)$.

The force of a particle can be computed in two ways.
The first is known as the \emph{ik-differentiation},
which takes the negative gradient w.r.t.~particle coordinate on the reciprocal
energy~\eqref{eqn:es-rec},
and then interpolates the complex exponential $e^{2\pi i\vect m\cdot\vect r}$.
This yields
\begin{align}\label{eqn:ik}
  \vect F^\ik_{\rec,i}
  \approx \ &
  q_i  
  \sum_{l_1,l_2,l_3}
  P_{\vect r_i} (l_1,l_2,l_3)
  [\,Q \ast (\vect G B^2)^{\vee}] (l_1,l_2,l_3),
\end{align}
where
\begin{align}
  \vect G(\vect m) = -4\pi i\vect m F(\vect m).
\end{align}
The alternative way is called the \emph{analytical differentiation},
which takes the negative gradient on the approximated
reciprocal energy~\eqref{eqn:approx-rec-ener}, and leads to
\begin{align}\label{eqn:ad}
  \vect F^\ad_{\rec,i}
  \approx
  q_i  
  \sum_{l_1,l_2,l_3}
  -2\nabla_{\vect r_i}P_{\vect r_i} (l_1,l_2,l_3)
  [\,Q \ast ( F B^2)^{\vee}] (l_1,l_2,l_3).
\end{align}

\noindent
\textbf{Remark}: The ik-differentiation needs four FFTs:
one forward for $\hat Q$, and three backwards for the three
components of $\hat Q (\vect G B^2)$. The three backward
FFTs are independent and of the same size,
so they can be executed as a multiple variable FFT, which
usually saves time comparing with executing three single variable FFTs
consecutively.

\section{Multiple staggered mesh Ewald method}
\label{sec:msme}

\begin{figure}
  \centering
  \includegraphics[width=0.49\textwidth] {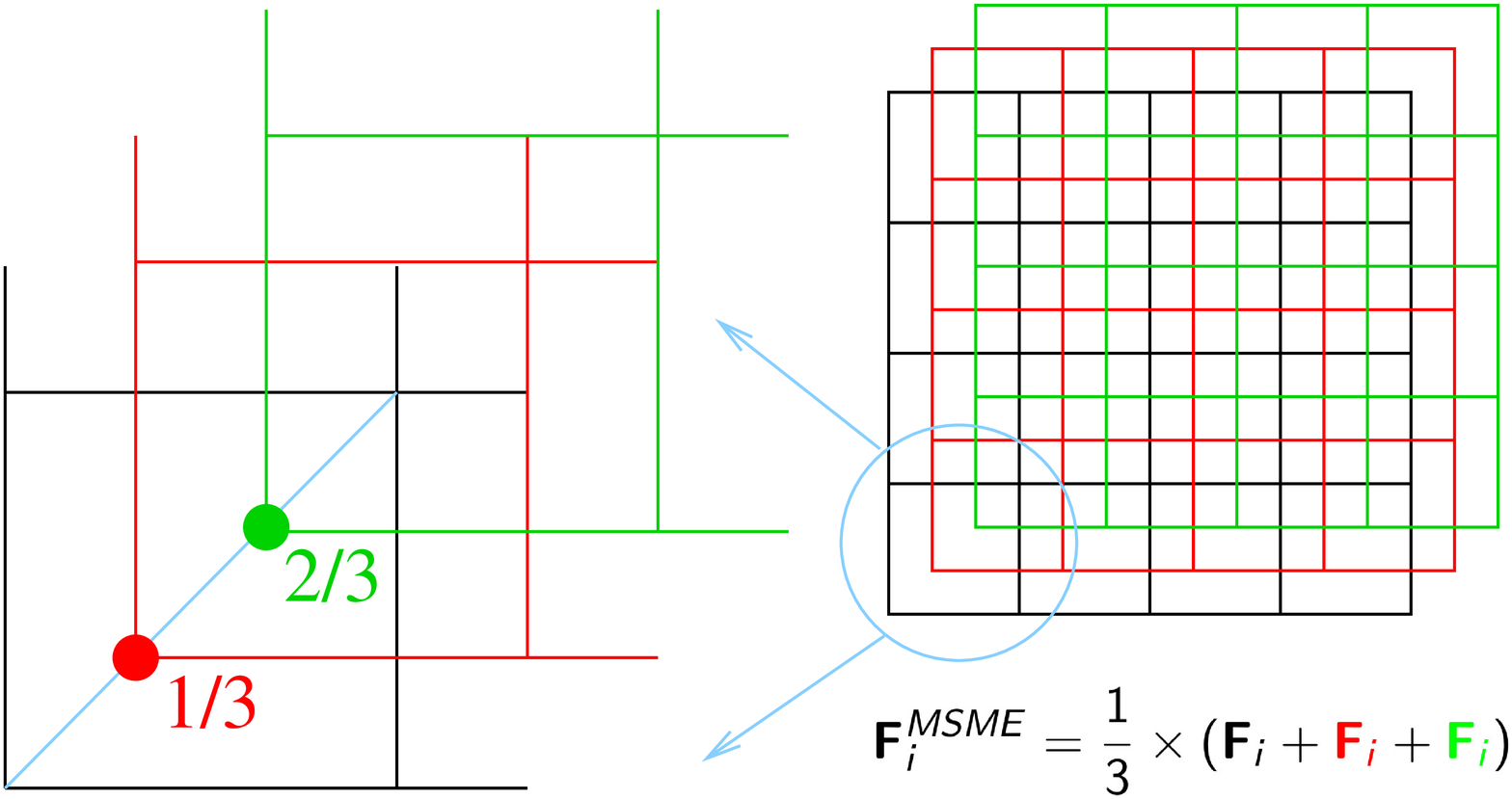}
  \caption{The schematic plot of a triple staggered mesh Ewald.
    The $\vect F_i$ with different color on the r.h.s.~of the equation denotes the reciprocal SPME force computed on the mesh with the same color.
  }
  \label{fig:ms}
\end{figure}

The MSME method averages the
reciprocal interactions computed by SPME on $M$ identical meshes,
which locate on the $M$ equally partitioning points of
the mesh subcell diagonal,
as illustrated by Fig.~\ref{fig:ms}.
The reciprocal force can be computed by either ik- or
analytical differentiation,
but all meshes in MSME should use the identical force
scheme.
These two branches of MSME are named by IK-MSME and AD-MSME, respectively.
The MSME method can be easily implemented by using the existing
SPME codes, because the computation on each mesh is the standard SPME and
the different meshes are independent.

We claim that the reciprocal accuracy of MSME that uses $M$ meshes is either of the following two cases:
\begin{enumerate}[label=\Roman*.]
\item
  In a certain range of parameter $\beta$,
  the accuracy is the same as the SPME that refines the mesh on each direction by $M$ times,
  i.e.~using an $MK_1 \times MK_2 \times MK_3$ mesh.
\item
  In the complementary range of $\beta$,
  the accuracy is almost the same as
  the SPME that uses twice order of the B-spline interpolation ($2n$ in our notation).
\end{enumerate}
These claims will be carefully checked both by numerical examples and by theoretical
accuracy analyses.



The numerical investigation of the MSME is carried out in two testing systems:
\begin{itemize}
\item \textbf{System 1}:
  The simulation region is a cubic cell of size
  $3.724\textrm{nm} \times 3.724\textrm{nm} \times 3.724\textrm{nm}$.
  In total, 5184 charged particles are uniformly and independently distributed
  in the region.
  Each of the 1728 particles carries a negative partial charge of $-0.834\,e$,
  while each of the rest (3456) particles carries a positive partial charge of $0.417\,e$.
  The total charge of the system is neutral.
\item \textbf{System 2}:
  The simulation region is a cubic cell of size 
  $3.724\textrm{nm} \times 3.724\textrm{nm} \times 3.724\textrm{nm}$,
  and contains 1728 TIP3P~\cite{jorgensen1983comparison} water molecules.
  The partial charges on the hydrogen and oxygen atoms are $0.417\,e$ and $-0.834\,e$, respectively.
  The configuration was taken from an equilibrium NPT simulation~\cite{gao2016sampling} at 300~K and 1~Bar.
  It should be noted that the number of particles and
  the amount of the partial charge of each particle are exactly the same
  as the System~1.
  The difference is that the charges in System 1
  are randomly and independently distributed,
  while the charges in System 2 are bonded and correlated
  according to the equilibrium water system configuration.
\end{itemize}
{All the numerical studies in this work are carried out by using our in-house MD package MOASP.}

The accuracy in computing the electrostatic interactions is measured by
the root means square (RMS) force error
that is defined by
\begin{align}\label{eqn:rms-ef}
  \me = \sqrt{
  \langle
  \vert \Delta \vect F\vert^2
  \rangle
  },
  \quad
  \Delta \vect F = \vect F - \vect F^\ast,
\end{align}
where
$\vect F$
and
$\vect F^\ast$
are the computed and accurate electrostatic forces, respectively.
The RMS force error will be called the ``error'' in short. 
The direct and reciprocal errors are defined similarly by
replacing the electrostatic force in Eq.~\eqref{eqn:rms-ef} with
the direct and reciprocal forces, respectively.
The direct force of MSME is computed in the same way as the Ewald summation,
the accuracy of which has already been studied~\cite{kolafa1992cutoff,wang2012error}.
Therefore, if not stated otherwise, the error estimate is 
developed only for the reciprocal error.

\begin{figure}
  \centering
  \includegraphics[width=0.45\textwidth]{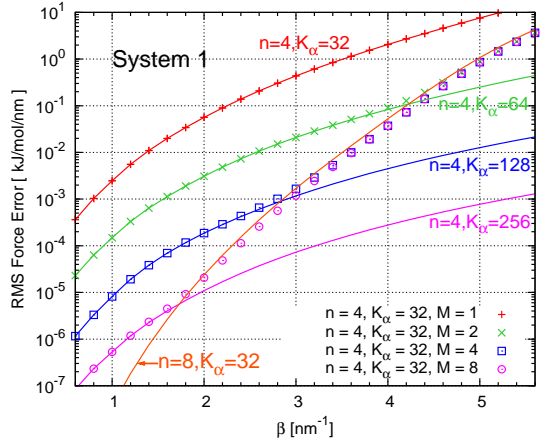}
  \caption{The reciprocal error of the IK-MSME as a function of the
    splitting parameter $\beta$ investigated in System 1.
    The chromatic solid line
    are errors
    computed from the IK-SPME method (parameters presented along with the lines),
    while
    the red ``$+$'',
    the green ``$\times$'',
    the blue ``$\boxdot$'' and
    the pink ``$\odot$''
    denote the reciprocal errors    
    computed from the IK-MSME
    with $M= 1, \ 2, \ 4$ and 8, respectively.
    The $M=1$ IK-MSME is identical to the IK-SPME method.
  }
  \label{fig:ik-sys1}
\end{figure}

\begin{figure}
  \centering
  \includegraphics[width=0.45\textwidth]{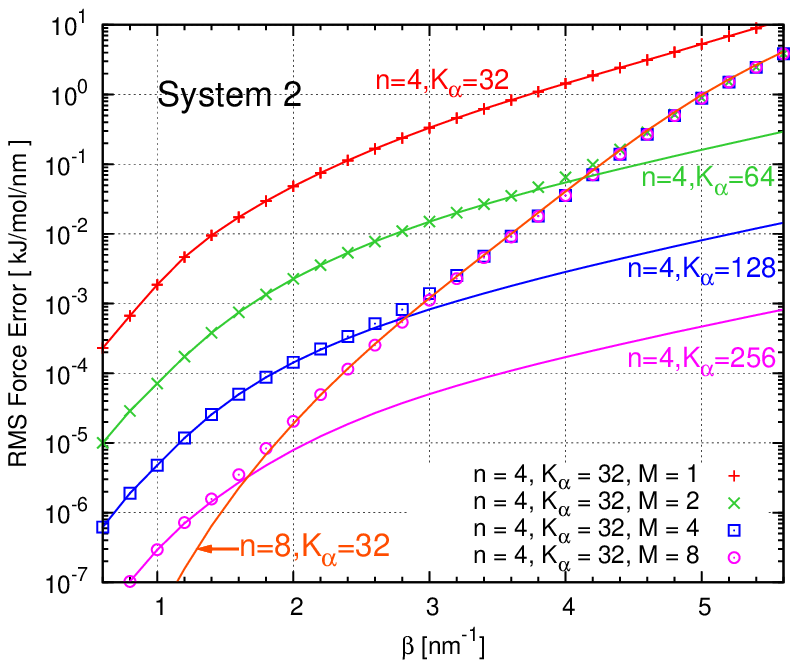}
  \caption{The reciprocal error of the IK-MSME as a function of the
    splitting parameter $\beta$ investigated in System 2.
    The meaning of the symbols is the same as Fig.~\ref{fig:ik-sys1}.
  }
  \label{fig:ik-sys2}
\end{figure}

\begin{figure}
  \centering
  \includegraphics[width=0.45\textwidth]{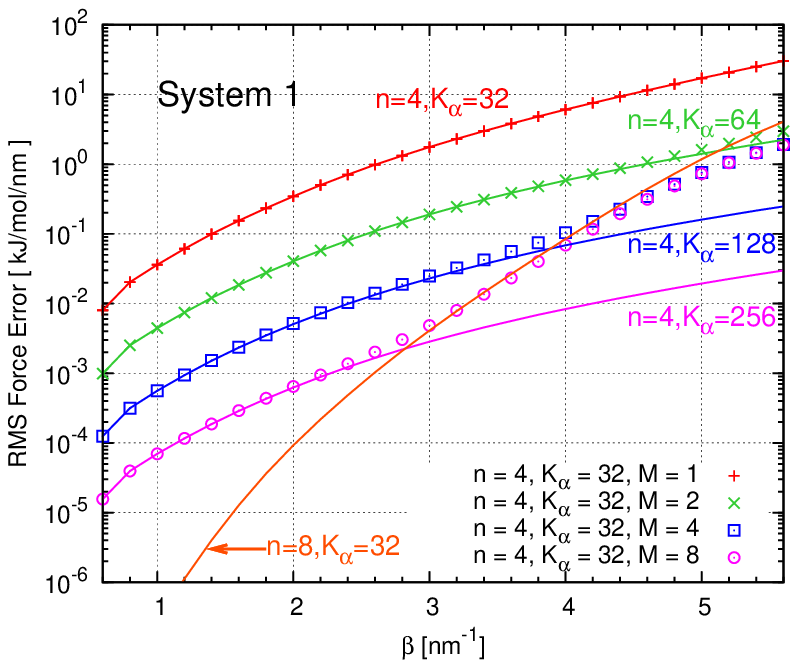}
  \caption{The reciprocal error of the AD-MSME as a function of the
    splitting parameter $\beta$ investigated in System 1.
    The chromatic solid line
    are errors
    computed from the AD-SPME method (parameters presented along with the lines),
    while
    the red ``$+$'',
    the green ``$\times$'',
    the blue ``$\boxdot$'' and
    the pink ``$\odot$''
    denote the reciprocal errors    
    computed from the AD-MSME
    with $M= 1, \ 2, \ 4$ and 8, respectively.
    The $M=1$ AD-MSME is identical to the  AD-SPME method.
  }
  \label{fig:ad-sys1}
\end{figure}

\begin{figure}
  \centering
  \includegraphics[width=0.45\textwidth]{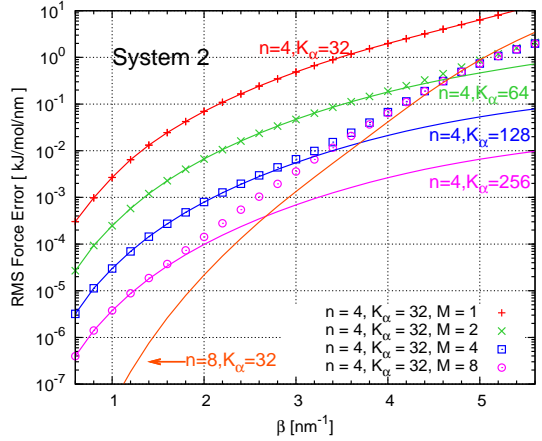}
  \caption{The reciprocal error of the AD-MSME as a function of the
    splitting parameter $\beta$ investigated in System 2.
    The meaning of the symbols is the same as Fig.~\ref{fig:ad-sys1}.
  }
  \label{fig:ad-sys2}
\end{figure}

The reciprocal errors of the IK-MSME in Systems 1 and 2 are presented
in Figs.~\ref{fig:ik-sys1} and \ref{fig:ik-sys2}, respectively.
The reciprocal errors of AD-MSME in Systems 1 and 2 are presented in Figs.~\ref{fig:ad-sys1} and \ref{fig:ad-sys2}, respectively.
For both force schemes, we investigate the number of
meshes $M=1, \ 2, \ 4, \ 8$ 
with the interpolation order $n=4$ and the number of mesh points $K_1 = K_2 = K_3 = 32$.
As a comparison, we present the reciprocal error of the SPME
that uses $n=8$, $K_\alpha = 32$ and $n=4$, $K_\alpha = 32, \ 64, \ 128, \ 256$.
The $M=1$ MSME is identical to the SPME that uses the same mesh ($K_\alpha=32$) and interpolation order ($n=4$).
The accurate forces are computed by well-converged Ewald summations. 
In the Figures,
the points present the reciprocal error of
the MSME method, 
and the solid lines present the reciprocal error of 
the SPME method.
The numerical phenomena of both force schemes in both systems are
similar:
At relatively small $\beta$, the accuracy of
the MSME that uses $M$ meshes matches the SPME that uses a mesh refined $M$ times on three directions,
while at relatively large $\beta$, the  MSME error of $M\geq 2$ roughly follows the error of SPME that uses twice of the
B-spline interpolation order.
These numerical results are clear evidences that support our claims I and II.


\section{Discussion of the numerical phenomena and the error estimate }
\label{sec:error}

The numerical phenomena of the MSME observed in the testing systems are understood
by the error estimate, which is derived under the framework
proposed in Ref.~\cite{wang2012numerical}.
It has been shown that the error
of a pairwise interaction (electrostatic interaction is pairwise) is composed of three additive parts, which are
the homogeneity, inhomogeneity and correlation errors
\begin{align}
  \me^2 = \me^2_\homo + \me^2_\hetero + \me_\correlation.
\end{align}
The homogeneity error stems from the fluctuation of the error force
$\Delta\vect F$ (see Eq.~\eqref{eqn:rms-ef}), while 
the inhomogeneity error is due to the inhomogeneous
charge distribution and is essentially the bias of $\Delta\vect F$.
If the positions of the charges in the system are correlated
(due to
e.g.~covalent bonds, hydrogen bonds, van der Waals interaction and so on), then
the correlation error arises.
The homogeneity and inhomogeneity error contributions are positive definite,
while the correlation error contribution may be negative,
which means that the charge correlation reduces the error in force computation~\cite{wang2012numerical}.

In this work, for simplicity, we assume that the charges are uniformly distributed,
so the inhomogeneity error vanishes.
The error estimate that only includes the homogeneity error is precise for the systems,
in which the charges are uniformly and independently distributed (e.g.~System 1).
Besides the homogeneity error,
we estimate the correlation error by using the nearest neighbor approximation technique~\cite{wang2012numerical}
for System 2 (TIP3P water).
This correlation error estimate can be directly extended
to other rigid water models containing three point charges, like
the SPC~\cite{berendsen1981interaction}, SPCE~\cite{berendsen1987missing} and TIP4P~\cite{jorgensen1983comparison} water models.
The error estimate is provided without proof, and the readers are referred to
the supplementary material 
for more details on the derivation of the error estimate.
{
  The supplementary material also provides the error kernels for the IK- and AD-MSME,
  from which the error estimates can be easily extended to systems that have non-uniform charge distributions.
}



\subsection{ik-differentiation}

Firstly we introduce the short-hand notation:
\begin{align}
  \za m = \frac{\hat \myphi(m_\alpha + l K_\alpha)}{\hat \myphi(m_\alpha)},
\end{align}
By using the definition of $\hat \myphi$, we have the estimate of
$ \vert \za m\vert \leq
  \vert
   {m_\alpha}/ {(m_\alpha + lK_\alpha)}
   \vert^n
$.
Noticing that $-K_\alpha/2 \leq m_\alpha < K_\alpha / 2$ and
$l\neq 0$, $\vert \za m\vert $ is fast decaying w.r.t.~growing $\vert l\vert$.
We further denote
\begin{align}
  \mg_{\alpha, l}(\vect m) &= \vect G(\vect m) \za m, \\
  \mg_{\alpha,l_1;\beta,l_2} (\vect m) &= \vect G(\vect m) \zaa m \zbb m.
\end{align}
$\mg_{\alpha, l}(\vect m)$ has the first powers of $\za m$,
and $\mg_{\alpha,l_1;\beta,l_2} (\vect m)$
has the second powers of $\za m$.
Therefore, in general, we have $\vert \mga m \vert \gg \vert \mgab m\vert $ for $l\neq 0$.
The homogeneity error is estimated, for IK-MSME, by (see Eqs.~(S52)--(S54))
\begin{align} \label{eqn:error-ik-homo}
  \vert \me^\ik_\homo \vert ^2
  \approx \,&
  \vert \me^{\ik,(1)}_\homo \vert ^2 +
  \vert \me^{\ik,(2)}_\homo \vert ^2.
\end{align}
On the r.h.s., $\me^{\ik,(1)}_\homo $ and
$ \me^{\ik,(2)}_\homo$ denote the
first and second order homogeneity errors, respectively,
and are defined by
\begin{align}\label{eqn:error-ik-homo-o1}
  \vert \me^{\ik,(1)}_\homo \vert ^2
  = &\,
  2q^2Q^2
  \sum_\alpha\sum_{l\neq 0}
  \sum_{\vect m}
  \theta_M(l)
  \vert
  \mg_{\alpha,l}(\vect m)
  \vert^2, \\ \nonumber
  \vert \me^{\ik,(2)}_\homo \vert ^2
  = &\,
  q^2Q^2
  \bigg\{
  \sum_{\alpha\neq\beta}\sumlab
  \sum_{\vect m}
  \theta_M(l_1 + l_2)
  \vert
  \mg_{\alpha,l_1;\beta,l_2}(\vect m)
  \vert^2
  \\\label{eqn:error-ik-homo-o2}
  & \quad
  +
  \sum_{\alpha,\beta}\sumlab
  \sum_{\vect m}
  \theta_M(l_1 - l_2)
  \vert
  \mg_{\alpha,l_1;\beta,l_2}(\vect m)
  \vert^2        
  \bigg\},  
\end{align}
where $Q^2 = \sum_i q_i^2$, $q^2 = Q^2/N$.
In the estimates~\eqref{eqn:error-ik-homo-o1} and \eqref{eqn:error-ik-homo-o2},
the function $\theta_M$ is defined by
\begin{align}
  \theta_M(l) =
  \begin{cases}
    0 & \quad (l\bmod M)\neq 0, \\
    1 & \quad (l\bmod M) = 0,
  \end{cases}
\end{align}
and is introduced by using the multiple staggered meshes.
For the IK-MSME with $M=1$ and $\theta_M(l) \equiv 1$,
noticing $\vert \mga m \vert \gg \vert \mgab m\vert $,
we have $\vert \me^{\ik,(1)}_\homo \vert \gg
\vert \me^{\ik,(2)}_\homo \vert
$, thus
the first order error dominates the reciprocal error.
This is why the error estimate of SPME ($M=1$ MSME)
only considered the first order contribution~\cite{wang2010optimizing}.

In the cases of non-trivial IK-MSME where $M \geq 2$,
we compute the first order error~\eqref{eqn:error-ik-homo-o1} by
\begin{align}\nonumber
  &\vert \me^{\ik,(1)}_\homo (\beta, n, \vect K, M) \vert ^2 \\\nonumber
  & =
    2q^2Q^2
    \sum_{\alpha}\sum_{l\neq 0}\sum_{m}
    \theta_M(l) \vert \mg_{\alpha,l}(\vect m) \vert^2 \\ \nonumber
  & =
    2 q^2Q^2
    \sum_{\alpha}\sum_{l\neq 0}\sum_{m}
    \vert \mg_{\alpha,Ml}(\vect m) \vert^2 \\\nonumber
  &=
    2 q^2Q^2
    \sum_{\alpha}\sum_{l\neq 0}\sum_{m}
    \Big\vert
    \vect G(\vect m)
    \frac{\hmyphi (m_\alpha + MlK_\alpha) }{ \hmyphi (m_\alpha) }
    \Big\vert^2 \\\label{eqn:1st-ik-identity}
  &=
  \vert \me^{\ik,(1)}_\homo (\beta, n, M\vect K, 1) \vert ^2.
\end{align}
Here we explicitly write the dependency of the error estimate
on the working parameters: $\beta$ (splitting parameter), $n$ (B-spline order),
$\vect K = (K_1,K_2,K_3)$ (number of mesh points) and  $M$ (number of meshes).
The last equation holds because of the identity $MlK_\alpha = l (MK_\alpha)$.
Thus the leading order error of MSME using $M$ identical $K_1\times K_2\times K_3$ meshes is
the same as the IK-SPME using an $MK_1\times MK_2\times MK_3$ mesh.

In the estimate of the second order error~\eqref{eqn:error-ik-homo-o2},
only the terms with $\alpha = \beta$ contribute significantly (will be shown by numerical examples later), and in this case $l_1 \neq l_2$ terms are much smaller than the $l_1 = l_2$ terms
due to the fast decaying of $\za m$ w.r.t.~$\vert l\vert$.
Therefore, we have
\begin{align} \nonumber
  & \vert \me^{\ik,(2)}_\homo (\beta, n, \vect K, M) \vert ^2 \\\nonumber
  & \approx
    \sum_{\alpha}\sum_{l\neq 0}\sum_{m}
    \vert \boldsymbol{\mg}_{\alpha,l;\alpha,l} (\vect m)\vert ^2 \\ \nonumber
  &=
      q^2Q^2
    \sum_{\alpha}\sum_{l\neq 0}  \sum_{m}
    \Big\vert
    \vect G(\vect m)
    \Big[
    \frac{\hmyphi (m_\alpha + lK_\alpha) }{ \hmyphi (m_\alpha) }
    \Big]^2
    \Big\vert ^2 \\ \nonumber
  &= 
      q^2Q^2
    \sum_{\alpha}\sum_{l\neq 0}  \sum_{m}
    \Big\vert
    \vect G(\vect m)
    \frac{\hat{\varphi}_{2n} (m_\alpha + lK_\alpha) }{ \hat{\varphi}_{2n} (m_\alpha) }
    \Big\vert ^2 \\\label{eqn:2nd-ik-identity}
  &= 
    \frac12 \,\vert \me^{\ik,(1)}_\homo (\beta, 2n, \vect K, 1) \vert^2.
\end{align}
The second equation holds because of the definition of $\hmyphi(m)$, i.e.~Eq.~\eqref{eqn:hat-phi}.
Eq.~\eqref{eqn:2nd-ik-identity} means that the second order error of the IK-MSME
is approximately $1/\sqrt 2$ times of the first order error of the IK-SPME method using twice order of the B-spline interpolation.

As the number of staggered meshes $M$ increases, the first order error decreases,
while the second order error does not.
Therefore, the Claim II (see Sec.~\ref{sec:msme}) holds
in the $\beta$ range that the first order error decreases to smaller than the
second order error.
The Claim I holds in the complementary range, in which the first order error still dominates.
The boundary of the ranges can be determined by solving the equation
$\me^{\ik,(1)}_\homo (\beta, n, \vect K, M) = \me^{\ik,(2)}_\homo (\beta, n, \vect K, M)$
w.r.t.~variable $\beta$.

The correlation error in the water system
contains bonded and non-bonded contributions.
The former is easily estimated by the nearest neighbor approximation with the knowledge of the
O-H bond length and H-O-H angle, while
the latter is partially  estimated by the same technique with the knowledge of
the radial distribution functions (RDFs).
It should be noted that the estimate for the bonded correlation error
is \emph{a priori}, because the values of O-H bond length and H-O-H angle are set up
by the water model, and are constrained over the simulations
(for the rigid water models).
By contrast,
the estimate for the non-bonded correlation error is \emph{a posteriori},
because
the RDFs are, in general, computed out of the molecular configurations sampled by the simulation.
Ref.~\cite{wang2012numerical} showed that, for a water system,
the bonded contribution
accounts for the substantial part of the correlation error,
and is good enough for the applications like the parameter tuning~\cite{wang2012numerical}.
Therefore, we only consider the nearest neighbor approximation for the bonded correlation error.

For the MSME using the ik-differentiation,
the nearest neighbor approximation to the
correlation error in a three-point-charge water system is (see Eq.~(S65))
\begin{align} \nonumber
  \me^\ik_\correlation = &\,
  q^2 Q^2
  \suma \sumll
  \sum_{m}
  \theta_M(l) T^w(\vect m) \vert \mg_{\alpha,l}(\vect m) \vert^2  \\
  &+
  q^2 Q^2
  \suma \sumll
  \sum_{m}
  \theta_M(l) T^w(\vect m + lK_\alpha \vect a_\alpha^\ast) \vert \mg_{\alpha,l}(\vect m) \vert^2,
\end{align}
where the function $T^w$ provides the information of the bonded charge correlation in the water molecule, and is defined by
\begin{align}
  T^w(\vect m)
  & =
  \frac{4\qh \qo}{2\qh^2 + \qo^2} T_{ \so}(\vect m)
  +
  \frac{2\qh^2}{2\qh^2 + \qo^2} T_{\sh}(\vect m).
\end{align}
$\so$ is the vector connecting the oxygen and the hydrogen atoms, and
$\sh$ is the vector connecting two hydrogen atoms. The function $T_{\vect b}(\vect m)$ is the structure factor of bond $\vect b$ averaged over all possible directions,
therefore, it only depends on the size of the Fourier mode $m = \vert \vect m\vert$, and the bond length $b = \vert \vect b\vert$, and writes
\begin{align}\label{eqn:tb-sin}
  T_{\vect b}(\vect m)
  = \langle e^{2\pi i \vect m\cdot \vect b}\rangle_{\mathrm{directions}}
  =
  \frac{\sin(2\pi mb)}
  {2\pi mb}.
\end{align}
In the ensemble average in Eq.~\eqref{eqn:tb-sin}, we assumed that all directions are equally
possible.
For TIP3P water model, $\vert \so \vert = 0.09572$~nm, and
$\vert \sh\vert = 0.15139$~nm.

\subsection{Analytical differentiation}

An extra error due to the ``self-interaction''
presents in the AD-SPME force computation, 
and can be removed by subtracting the analytic formula of the
self-interaction from the force~\cite{ballenegger2011removal,wang2012numerical}.
For the AD-MSME, 
the self-interaction is given by (see Eq.~(S79))
\begin{align} \nonumber
  \vect F^\self_i
  = &\,
      q_i^2
      \sum_{\alpha}\sum_{l\neq 0}
      \sum_{\vect m}
      \theta_M(l)
      \mf_{\alpha,\l}(\vect m)\,
      e^{2\pi i l u_\alpha}  \\ \nonumber
    & +
      q_i^2
      \sum_{\alpha\neq \beta}\sumlab
      \sum_{\vect m}
      \theta_M(l_1 + l_2)\,
      \mf_{\alpha,l_1;\beta,l_2}(\vect m)\,
      e^{2\pi i (l_1 u_\alpha + l_2 u_\beta)} \\
    & +
      q_i^2
      \sum_{\alpha, \beta}\sumlab
      \sum_{\vect m}
      \theta_M(l_1 - l_2)\,
      \mf_{\alpha,l_1;\beta,l_2}(\vect m)\,
      \,e^{2\pi i (l_1 u_\alpha - l_2 u_\beta)},
\end{align}
where we introduce the notations
\begin{align}
  \mfa m
  &=
  -4\pi i l K_\alpha \vect a^\ast_\alpha F(\vect m)  Z_{\alpha,l}(\vect m), \\
  \mfab m
  &=
  -4\pi i l_1 K_\alpha \vect a^\ast_\alpha F(\vect m)  Z_{\alpha,l_1}(\vect m) Z_{\beta,l_2}(\vect m).
\end{align}
We also have $\vert \mfa m \vert \gg \vert \mfab m\vert $ for $l\neq 0$.
When the self-interaction is removed,
the homogeneity error of the AD-MSME is estimated
as (see Eqs.~(S93)--(S95))
\begin{align} 
  \vert \me^\ad_\homo\vert ^2
  \approx 
  \vert \me^{\ad,(1)}_\homo \vert ^2 +
  \vert \me^{\ad,(2)}_\homo \vert ^2,
\end{align}
where, similar to the IK-MSME, the homogeneity error of the AD-MSME is also composed
of the first and second order contributions, which are defined by
\begin{align} \label{eqn:error-ad-homo-o1}
  \vert \me^{\ad,(1)}_\homo\vert ^2
  = &\,
      q^2Q^2
      \sum_{\alpha}\sum_{l\neq 0}
      \sum_{m}
      \theta_M(l)
      \Big[
      \vert \mg_{\alpha,l}(\vect m)
      \vert^2
      +
      \vert \mg_{\alpha,l}(\vect m) + \mf_{\alpha,l}(\vect m) \vert^2
      \Big].
  \\ \nonumber
  \vert \me^{\ad,(2)}_\homo\vert^2
  = &\,
      q^2Q^2
      \bigg\{
      \sum_{\alpha\neq\beta}\sum_{l_1,l_2\neq 0}
      \sum_{m}
      \theta_M(l_1+l_2)
      \Big[\,
      \frac12\, \vert \mg_{\alpha,l_1;\beta,l_2}(\vect m) \vert^2 \\\nonumber
    & \qquad+
      \vert \frac12\, \mg_{\alpha,l_1;\beta,l_2}(\vect m) + \mf_{\alpha,l_1;\beta,l_2}(\vect m) \vert^2 \\ \nonumber
    & \qquad+
      \Big(
      \frac12 \mg_{\alpha,l_1;\beta,l_2}(\vect m) + \mf_{\alpha,l_1;\beta,l_2}(\vect m)
      \Big)
      \cdot
      \Big(
      \frac12 \mg_{\alpha,l_1;\beta,l_2}(\vect m) + \mf_{\beta,l_2;\alpha,l_1}(\vect m)
      \Big)
      \Big]
  \\\label{eqn:error-ad-homo-o2}
    & +
      \sum_{\alpha,\beta}\sum_{l_1,l_2\neq 0}
      \sum_{m}
      \theta_M(l_1-l_2)
      \vert  \mg_{\alpha,l_1;\beta,l_2}(\vect m) + \mf_{\alpha,l_1;\beta,l_2}(\vect m) \vert^2
      \bigg\}.
\end{align}
The first order error in the estimate is
\begin{align} \nonumber
  &\vert \me^{\ad,(1)}_\homo (\beta, n, \vect K, M) \vert ^2 \\\nonumber
  &= 
    q^2Q^2
    \sum_{\alpha}\sum_{l\neq 0}
    \sum_{m}
    \theta_M(l)
    \Big[
    \vert \mg_{\alpha,l}(\vect m)
    \vert^2
    +
    \vert \mg_{\alpha,l}(\vect m) + \mf_{\alpha,l}(\vect m) \vert^2
    \Big] \\\nonumber
  &= 
    q^2Q^2
    \sum_{\alpha}\sum_{l\neq 0}
    \sum_{m}
    \Big[
    \vert \mg_{\alpha,Ml}(\vect m)
    \vert^2
    +
    \vert \mg_{\alpha,Ml}(\vect m) + \mf_{\alpha,Ml}(\vect m) \vert^2
    \Big] \\\nonumber
  &= 
    q^2Q^2
    \sum_{\alpha}\sum_{l\neq 0}
    \sum_{m}
    \Big[
    \vert \vect G(\vect m) \vert^2
    +
    \vert \vect G(\vect m) - 4\pi i Ml K_\alpha\vect a^\ast_\alpha F(\vect m)\vert^2
    \Big] \times
    \Big\vert
    \frac{\hat{\varphi}_{n} (m_\alpha + MlK_\alpha) }{ \hat{\varphi}_{n} (m_\alpha) }
    \Big\vert ^2 \\
  & =
    \vert \me^{\ad,(1)}_\homo (\beta, n, M\vect K, 1) \vert ^2.
\end{align}
Thus the first order error of AD-MSME
using $M$ identical $K_1\times K_2\times K_3$ meshes is
the same as the AD-SPME using a mesh $M$ times finer on each direction, i.e.~an $MK_1\times MK_2\times MK_3$ mesh.

In the second order error, only the terms with $\alpha = \beta$, $l_1 = l_2$ contribute significantly,
so we have the approximation
\begin{align} \nonumber
  &\vert \me^{\ad,(2)}_\homo (\beta, n, \vect K, M) \vert ^2 \\\nonumber
  &\approx
    \sum_{\alpha}\sum_{l\neq 0}\sum_{m}
    \vert \boldsymbol{\mg}_{\alpha,l;\alpha,l} (\vect m) +
    \boldsymbol{\mf}_{\alpha,l;\alpha,l} (\vect m)\vert ^2
  \\ \nonumber
  &=
      q^2Q^2
    \sum_{\alpha}\sum_{l\neq 0}  \sum_{m}
    \vert
    \vect G(\vect m) - 4\pi i l K_\alpha\vect a^\ast_\alpha F(\vect m)
    \vert ^2
    \Big[
    \frac{\hmyphi (m_\alpha + lK_\alpha) }{ \hmyphi (m_\alpha) }
    \Big]^4
  \\ \nonumber
  &=
      q^2Q^2
    \sum_{\alpha}\sum_{l\neq 0}  \sum_{m}
    \vert
    \vect G(\vect m) - 4\pi i l K_\alpha\vect a^\ast_\alpha F(\vect m)
    \vert ^2
    \Big[
    \frac{\hat{\varphi}_{2n} (m_\alpha + lK_\alpha) }{ \hat{\varphi}_{2n} (m_\alpha) }
    \Big]^2
  \\  \nonumber
  &= 
    \vert \me^{\ad,(1)}_\homo (\beta, 2n, \vect K, 1) \vert ^2 - \frac12 \,\vert \me^{\ik,(1)}_\homo (\beta, 2n, \vect K, 1) \vert ^2
  \\\label{eqn:2nd-ad-identity}
  &\approx
    \vert \me^{\ad,(1)}_\homo (\beta, 2n, \vect K, 1) \vert ^2.
\end{align}
The last approximation holds because the IK-SPME is usually much more accuracy than the AD-SPME
when the parameters are the same (see, e.g.~Figs.~\ref{fig:ik-sys1} and \ref{fig:ad-sys1}).
Eq.~\eqref{eqn:2nd-ad-identity} means that the second order error of AD-MSME is
roughly the same as
the error of the AD-SPME with twice order
of the B-spline interpolation.

Finally, the nearest neighbor approximation to the
correlation error of
AD-MSME is given by (see Eq.~(S101))
\begin{align} \nonumber
  \me^\ad_\correlation
  = &\,
  q^2Q^2 
  \suma\sumll
  \sum_{m}
  \theta_M(l)\,
  T^w(\vect m) \vert \mg_{\alpha,l}(\vect m) + \mf_{\alpha,l}(\vect m)\vert^2\\
  & +
  q^2Q^2 
  \suma\sumll
  \sum_{m}
  \theta_M(l)\,
  T^w(\vect m + l K_\alpha\vect a_\alpha^\ast)
  \vert \mg_{\alpha,l}(\vect m) \vert^2.
\end{align}

\section{Numerical validation of the error estimate}
\label{sec:error-numeric}

\begin{figure}
  \centering
  \includegraphics[width=0.45\textwidth]{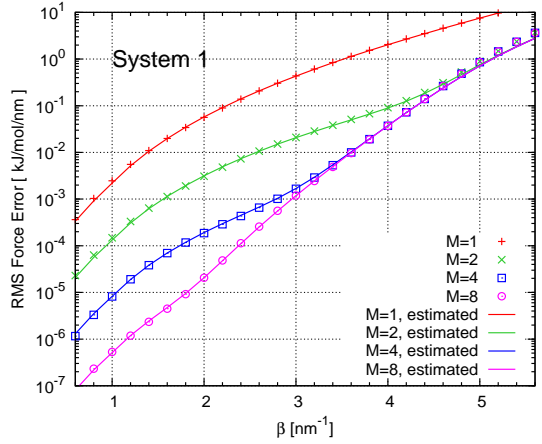}
  \caption{
    The reciprocal error (points) of the IK-MSME and the error estimate (solid lines)
    as a function of the
    splitting parameter $\beta$ investigated
    in System 1.
    The number of mesh points is $K_\alpha = 32$ and the order of B-spline interpolation
    is $n = 4$.
    The IK-MSME with $M=1,2,4$ and 8 are plotted with color red, green, blue and pink, respectively.
  }
  \label{fig:ik-esti-sys1} 
\end{figure}

\begin{figure}
  \centering
  \includegraphics[width=0.45\textwidth]{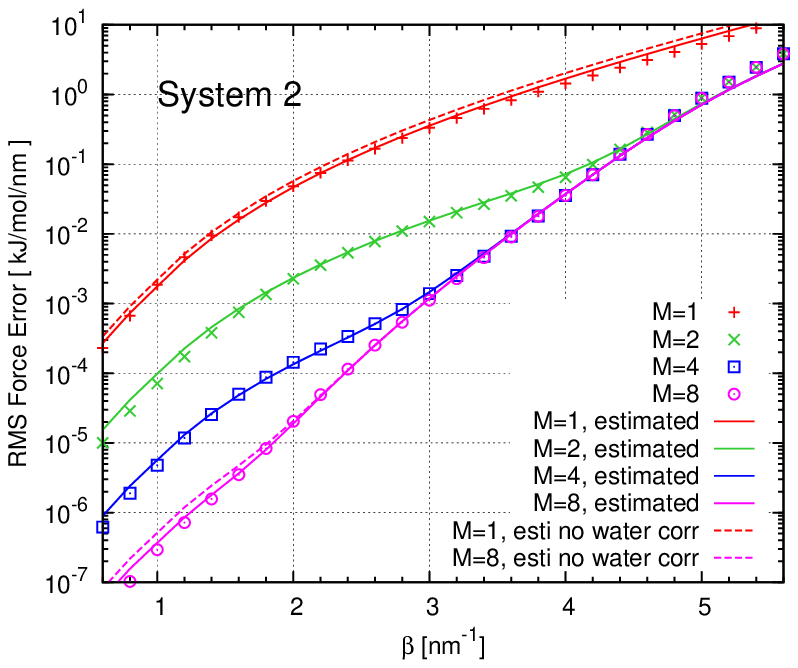}
  \caption{
    The reciprocal error (points) of the IK-MSME and its error estimate (solid lines)
    as a function of the
    splitting parameter $\beta$ investigated
    System 2 (bottom).
    The meaning of the symbols is the same as Fig.~\ref{fig:ik-esti-sys1}.
    The error estimate
    without taking into account the correlation error is shown by the dashed lines for $M=1$ and 8.
  }
  \label{fig:ik-esti-sys2} 
\end{figure}

\begin{figure}
  \centering
  \includegraphics[width=0.45\textwidth]{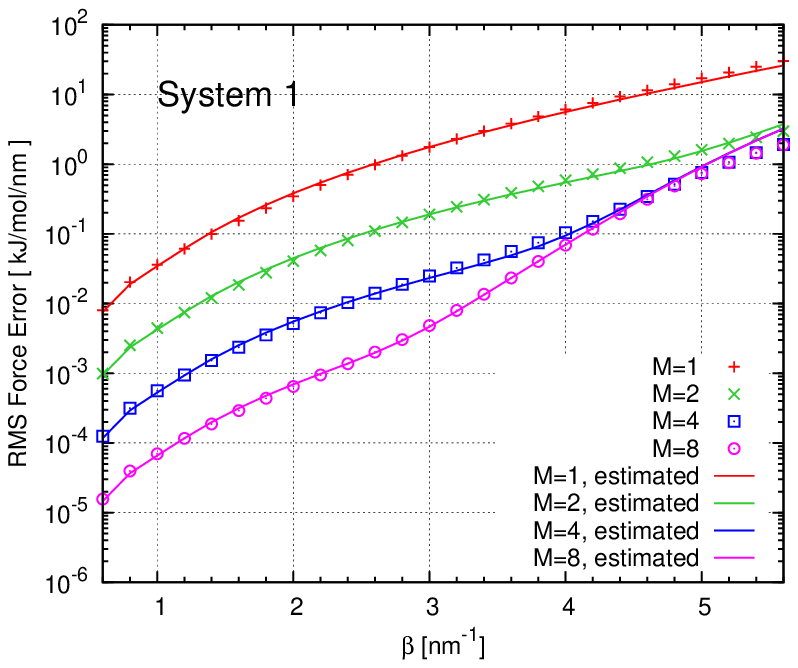}
  \caption{
    The reciprocal error (points) of the AD-MSME and the error estimate (lines)
    as a function of the
    splitting parameter $\beta$ investigated
    in System 1.
    The number of mesh points is $K_\alpha = 32$ and the order of B-spline interpolation
    is $n = 4$.
    The AD-MSME with $M=1,2,4$ and 8 are plotted with color red, green, blue and pink, respectively.
  }
  \label{fig:ad-esti-sys1} 
\end{figure}

\begin{figure}
  \centering
  \includegraphics[width=0.45\textwidth]{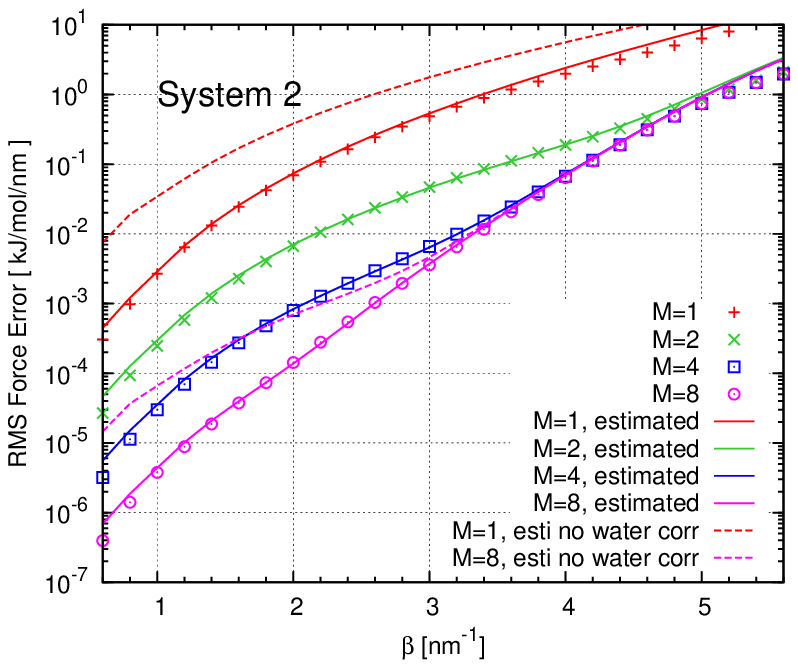}
  \caption{
    The reciprocal error (points) of the AD-MSME and its error estimate (lines)
    as a function of the
    splitting parameter $\beta$ investigated
    in System 2.
    The meaning of the symbols is the same as Fig.~\ref{fig:ad-esti-sys1}.
    The error estimate
    without taking into account the correlation error is shown by the dashed lines for $M=1$ and 8.  
  }
  \label{fig:ad-esti-sys2} 
\end{figure}

In this section we numerically check the quality of the error estimate.
The reciprocal error of the MSME
and the corresponding error estimate
are shown in Figs.~\ref{fig:ik-esti-sys1}--\ref{fig:ad-esti-sys2}.
The reciprocal error is obtained by comparing the MSME reciprocal force
with a well converged Ewald reciprocal force using the same splitting parameter $\beta$.
For System 1 (see Figs.~\ref{fig:ik-esti-sys1} and \ref{fig:ad-esti-sys1}), the error estimate is
sharp for both ik and analytical differentiations in most of the $\beta$ range.
The agreement between the actual and estimated errors is expected,
because the estimate catches all error contributions in a system that has a uniform and uncorrelated charge distribution.
{Deviation of the error estimate is observed when $\beta > 5.0\ \textrm{nm}^{-1}$, and the error in this range is larger than 1~kJ/mol/nm.
  The deviation may stem from the truncated terms in the Ewald summation, which become increasingly significant with larger parameter $\beta$~\cite{wang2010optimizing}.
  In this work, we do not consider this contribution, because the error estimate would become too complicated.
}



The error estimates of the IK- and AD-MSME in System 2
are presented and compared with the actual error in Figs.~\ref{fig:ik-esti-sys2} and \ref{fig:ad-esti-sys2}, respectively.
Both the error estimates of IK- and AD-MSME are sharp, 
but they are less precise when $\beta$ is smaller than $1.6\ \mathrm{nm}^{-1}$.
The maximum deviation at $\beta = 0.6\ \mathrm{nm}^{-1}$ is 87\%, 
which is acceptable for the applications like the parameter tuning~\cite{wang2010optimizing,wang2012numerical}.
The error estimates without counting the correlation error
are also presented in the Figures by the dashed lines.
For clarity, we only plot the $M=1$ and 8 cases.
It is noticed that the error of IK-MSME is not sensitive to the
charge correlation in the system,
while the error of the AD-MSME is very sensitive to the charge correlation:
At relatively small $\beta$, the charge correlation reduces the error by more than one order of magnitude.
This reminds us that, if the correlation error is important
but difficult to estimate,
simply using the error estimates without considering the charge correlation
may be more reliable for the IK-MSME than for the AD-MSME.

\begin{figure}
  \centering
  \includegraphics[width=0.45\textwidth]{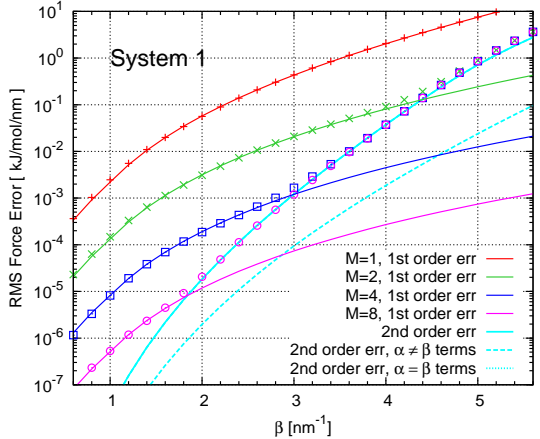}  
  \caption{The estimates of the first ($\me_\homo^{\ik,(1)}$) and second order ($\me_\homo^{\ik,(2)}$) homogeneity errors of the IK-MSME in System 1.
    The points plot the reciprocal errors of IK-MSME with
    $M=1$ (red ``$+$''),
    $2$ (green ``$\times$''),
    4 (blue ``$\boxdot$'') and
    8 (pink ``$\odot$'').
    The first order errors of $M=1$, 2, 4 and 8 are presented by red, gree, blue and pink lines, respectively.
    The second order error does not depends on $M$, therefore, only one line is plotted.
    The dashed cyan line plots the $\alpha \neq \beta$ contribution in the second order error.
    The dotted cyan line (overlapping with the solid line) plots the $\alpha = \beta$ contribution in the second order error.
  }
  \label{fig:ik-esti-comp}
\end{figure}

\begin{figure}
  \centering
  \includegraphics[width=0.45\textwidth]{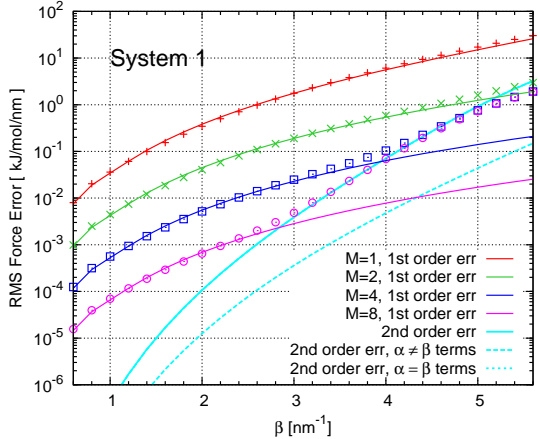}  
  \caption{
    The estimates of the first ($\me_\homo^{\ad,(1)}$) and second order ($\me_\homo^{\ad,(2)}$) homogeneity errors of AD-MSME in System 1.
    The meaning of the symbols is the same as Fig.~\ref{fig:ik-esti-comp}.
  }
  \label{fig:ad-esti-comp}
\end{figure}

{In Sec.~\ref{sec:error}, the Claims I and II were explained by 
the estimates of the first and second order homogeneity errors. 
We plot those of the IK- and AD-MSME for System 1 in Figs.~\ref{fig:ik-esti-comp} and \ref{fig:ad-esti-comp}, respectively.}
When $M=1$ (MSME reduces to the  SPME), the first order error dominates.
This explains why only the first order error was estimated for the SPME.
For the non-trivial MSME with $M > 1$, first order error dominates at the relatively small $\beta$, 
while the second order error dominates at the relatively large $\beta$.
{Therefore, we observe that the actual error firstly matches the first order error estimate at relatively small $\beta$,
and then follows the second order error estimate at relatively large $\beta$.}
When the $M$ increases, the first order error decreases, but the second order error does not,
therefore, the range that the first order error dominates shrinks.
The dashed and dotted cyan lines in Figs.~\ref{fig:ik-esti-comp} and \ref{fig:ad-esti-comp} denote the
$\alpha \neq\beta$  and $\alpha = \beta$ contributions (see Eqs.~\eqref{eqn:error-ik-homo-o2} and \eqref{eqn:error-ad-homo-o2})
in the second order error, respectively.
The $\alpha = \beta$ contribution overlaps with the second order error, while
the $\alpha \neq \beta$ contribution is at least one order of magnitude smaller, 
therefore, 
it is demonstrated that the $\alpha = \beta$ contribution dominates the second order error.

\section{Conclusion and discussion}
\label{sec:conclusion}

In this work, the multiple staggered mesh Ewald (MSME) method is
proposed to improve the accuracy of the smooth particle mesh Ewald (SPME) method.
It takes the average of the SPME reciprocal forces computed on $M$ staggered meshes,
and achieves, in a certain parameter range,  the same accuracy as the SPME that uses $M^3$ times more mesh points.
In the complementary parameter range, it is almost as accurate as the SPME that uses  twice order of the B-spline interpolation.
The reduction of the necessary FFT mesh points is
particularly interesting for the massively parallel computation of the electrostatic interaction,
because the FFT is a well known bottleneck in the parallel implementation due to the intensive all-to-all data communications among the processors.

The accuracy of the MSME is understood by a systematical error estimate.
We prove that
the multiple staggered meshes 
reduce the first order part of the error
as much as refining the FFT mesh in the SPME, and
does not change the second order part, which  is roughly the same
as doubling the interpolation order in the  SPME.
The error estimate and the theoretical analysis on different orders of the error
are validated by
both a uniform and uncorrelated charge system and a three-point-charge rigid water model.
The error estimate developed in this work is significant not only because
it explains the numerical phenomena of MSME,
but also because it is of key importance in the parameter tuning.

The difference between the $B$ function used by us (Eq.~\eqref{eqn:b})
and that proposed in the original SPME paper (Eq.~\eqref{eqn:b-orig})
should be noticed in the implementation of MSME.


The number of floating point operations of the MSME method (reciprocal space only) is $M$ times more than
the SPME method.
However, it does not mean that the time to solution is also $M$ times longer.
It is worth noting that the operations on the $M$ meshes are independent and identical (regardless of the constant shifts of the meshes),
thus, it is possible to utilize the single instruction multiple data (SIMD) architecture that is widely provided by the modern processors to
greatly reduce the time to solution.
For example, the Intel's AVX SIMD processes 4 floating point numbers of double precision at once,
which means that the ideal time to solution of an $M=4$ MSME can be as short as the SPME that does not use SIMD.
In this work,
we do not measure the wall execution time of the MSME, nor to compare it to the SPME.
One reason is that the current implementation of MSME in MOASP does not use the SIMD architecture,
therefore, the execution time would not reflect the highest performance one may obtain from MSME.
Secondly, the focus of this work is to introduce the method, and to provide the theoretical analysis on the accuracy.
The future work of MSME would be to optimize the implementation of the method, and to systematically test its performance on different hardware.




\section*{Acknowledgment}
The authors gratefully acknowledge the financial support from
National High Technology Research and Development Program of China under Grant 2015AA01A304.
X.G. is supported by the National Science Foundation of China under Grants 91430218 and 61300012.
H.W. is supported by the National Science Foundation of China under Grants 11501039 and 91530322.


\end{document}